\documentclass[conference]{IEEEtran}
\usepackage{graphicx}
\usepackage{epstopdf}
\graphicspath{{./images/}}
\begin{document}

\title{How Many Antennas Do We Need for \\Massive MIMO Channel Sounding? -\\Validating Through Measurement}

\author{
    \IEEEauthorblockN{Thomas~Choi\IEEEauthorrefmark{1}, Fran\c{c}ois~Rottenberg\IEEEauthorrefmark{1}, Peng Luo\IEEEauthorrefmark{1}, Jianzhong Zhang\IEEEauthorrefmark{2}, Andreas F. Molisch\IEEEauthorrefmark{1}}
    \IEEEauthorblockA{\IEEEauthorrefmark{1}University of Southern California, Los Angeles, CA, USA}
    \IEEEauthorblockA{\IEEEauthorrefmark{2}Samsung Research America, Richardson, TX, USA}
}
\maketitle

\begin{abstract}
This paper investigates the impact of the number of antennas  (8 to 64) and the array configuration on massive MIMO channel parameters estimation for multiple propagation scenarios at 3.5 GHz. Different measurement environments are artificially created by placing several reflectors and absorbers in an anechoic chamber. ``Ground truth'' channel parameters, \textit{e.g}, path angles, are obtained by geometry and trigonometric rules. Then, these are compared to the channel parameters ``extracted" by the applying Space-Alternating Generalized Expectation-Maximization (SAGE) algorithm on the measurements. Overall, the estimation errors for various array configurations and the multiple environments are compared. This paper will help to determine the appropriate configuration of the antenna array and the parameter extraction algorithm for outdoor massive MIMO channel sounding campaigns.
\end{abstract}

\linespread{0.85}
\section{Introduction}
\IEEEPARstart Massive MIMO, since its introduction, has been considered one of the key wireless technologies to increase channel capacity and coverage \cite{Marzetta}. Designing efficient and accurate massive MIMO channel sounders is of paramount importance to successfully characterize and evaluate its performance. Therefore, massive MIMO channel measurements must be conducted at various environments with a reliable channel sounder containing a well-calibrated array with a large number of antennas (usually $> 64$). Yet, massive MIMO channel sounding using a relatively low-cost switched array with a single receive radio frequency chain is challenging. On the one hand, a large amount of components have to be properly calibrated. On the other hand, the process of acquiring channel measurements represent a huge amount of data that has to be received sequentially in between one coherence time of the channel. The problem becomes even more severe when the channel is fast-varying with mobility. In such cases, measurements using only subset of antennas may be necessary.

\IEEEPARstart In this paper, the SAGE algorithm \cite{SAGE} is applied to extract channel parameters from channel measurements performed in several propagation environments in an anechoic chamber. Similar SAGE evaluation studies from chamber measurements have previously been conducted, including evaluations for Ultra Wideband channels \cite{UWB_SAGE} and for different types of 8-element circular arrays \cite{circular_SAGE}. In this paper, we focus on the accuracy of the algorithm when the number of antennas (8 to 64) and the number of paths (2 to 4) are varied. This will help 1) understand the trade-offs between antenna number and measurement duration and 2) select appropriate input parameters for SAGE depending on measurement environment.

\begin{figure}
    \includegraphics[scale=0.0623]{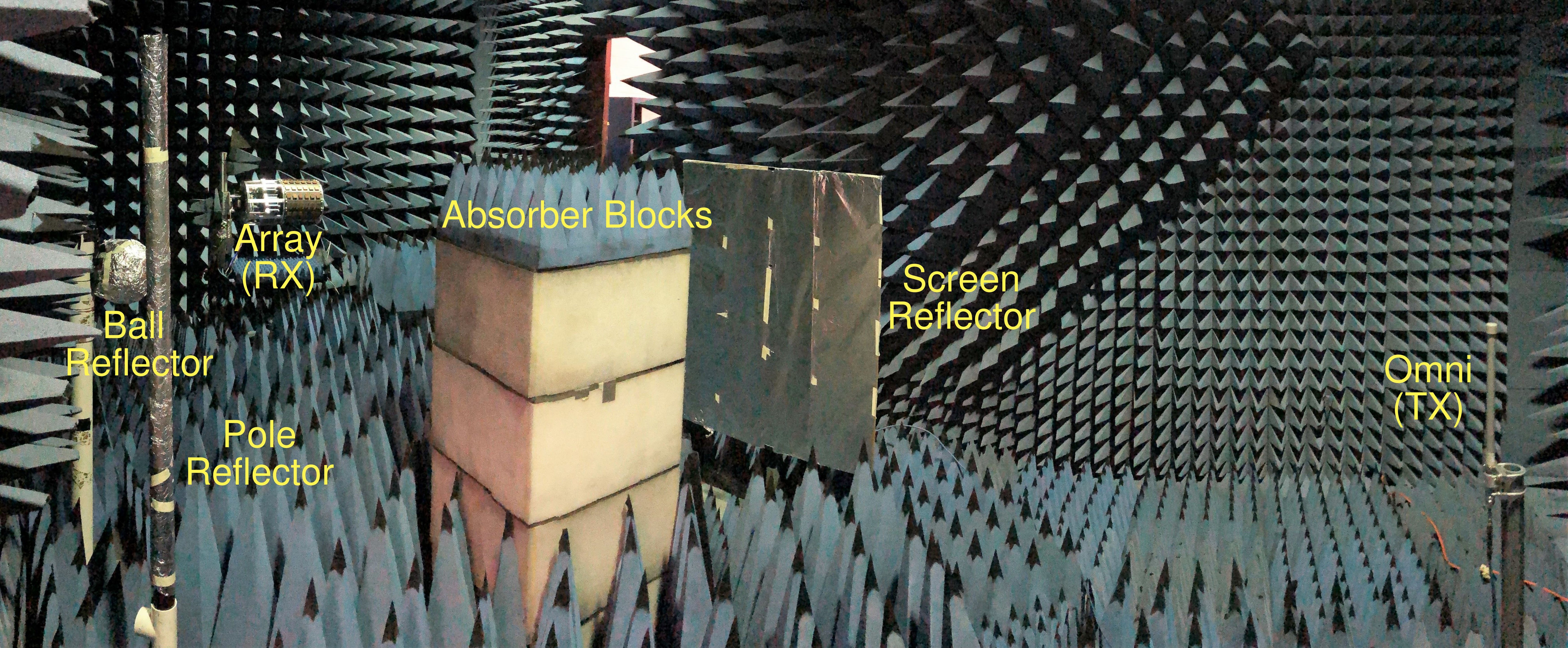}
    \caption{One of the anechoic chamber settings with an omni antenna, reflectors (ball, pole, and screen), absorber blocks, and a cylindrical array}
    \centering
    \label{fig:chamber}
	\vspace{-1.5em}
\end{figure}

\section{Measurement Setup}
\IEEEPARstart Fig. \ref{fig:chamber} shows one of the four measurement setups conducted within anechoic chamber at University of Southern California (USC). On the transmitting side, a high gain omni-directional antenna at 3.5 GHz is used. High gain was selected to reduce the beamwidth in the elevation and provide strongest beam at the center within elevation domain, which helps to avoid ambiguities of elevation angles in reflected paths. A center frequency of 3.5 GHz has been selected as it is directly foreseen for 5G wireless frequencies. 

\IEEEPARstart On the receive side, a cylindrical massive MIMO array based on parasitic patch antenna was used. The array contains 64 radiating patch elements (16 ``columns'' of 4x1 antenna elements, forming a 16-gon cylinder). Each patch element is soldered with two ports corresponding to two different polarizations (vertical/horizontal). Each antenna port has a bandwidth ($S11(f)$ (dB) $<$ -10 dB) of about 400 MHz from 3.3 to 3.7 GHz. The array is RF switch-based, sending or receiving signals from one port at a time. 

\IEEEPARstart Three types of reflectors with different shapes were used, including a screen, a ball, and a pole. The absorber blocks were also used to block the line-of-sight (LOS) depending on scenarios. Overall, the four scenarios considered were 1) screen and absorbers, 2) screen, pole, and absorbers, 3) screen, pole, ball, and absorbers, and 4) screen, pole, and ball.

\IEEEPARstart One important note is that the cylindrical array was ``lying down'' on a positioner. This implies that, from the point of array's view, the coordinate system (azimuth-elevation) as well as the polarization (vertical-horizontal) is rotated 90 degrees. In Section III., the angles will be discussed in  ``array's view''; e.g., in Fig. \ref{fig:chamber}, various reflected paths should differ only in elevation instead of azimuth, due to the rotation of coordinate system and low beamwidth (in azimuth from the array's view) of the omni-antenna. Because the straight-up omni-antenna is horizontally polarized in array's view, only 64 horizontally polarized ports were considered for this experiment.

\IEEEPARstart The measurements were done using a Vector Network Analyzer (VNA) switching across the receive elements. the relatively slow speed compared to time-domain setups was not an issue as the anechoic chamber is a static environment providing a coherent channel during the measurement. 401 frequency points across 400 MHz bandwidth (3.3 to 3.7 GHz) were used, and an amplifier was added before the VNA to increase the signal-to-noise ratio (SNR) to $>$ 50 dB.

\section{Channel Parameter Estimation Using SAGE}
\IEEEPARstart For each measurement, the ``ground truth'' channel parameters of three reflectors including delays, azimuth angles of arrival, and elevation angles of arrival were attained using a laser distance meter and simple trigonometric rules. These parameters were then compared to the channel parameters obtained from the SAGE. For the input of the SAGE, delay step size was selected to be 1/50 of the inverse of 400 MHz bandwidth, while angular step sizes were selected to be 1 degree for azimuth and 0.5 degree for elevation. The azimuth and the elevation angle search range have been limited to +/- 45 degrees from the LOS path with 0 degree azimuth and 90 degree elevation. The azimuth of all reflectors stayed at 0 degree while the elevation of the board, pole, and ball lied at 72, 113, and 122 degrees.

\IEEEPARstart The number of selected antenna ports ranged from 64 to 32 to 16 to 8, with 64 using all 16 columns separated by 22.5 degrees in azimuth, 32 containing 8 columns separated by 45 degrees, 16 containing 4 columns separated by 90 degrees, and 8 only containing 2 columns separated by 180 degrees. While there was only one estimation for 64 ports case, two estimations for 32 ports case were averaged, four estimations for 16 ports case were averaged and so on. The reason for selecting columns rather than rows was because elevation differed for reflectors from the array's view. Among many paths SAGE estimated, the path with closest elevation was selected, as long as the delay of the paths differed less than 1.5 ns (45 cm).

\begin{figure}
    \includegraphics[height=4cm, width=8.8cm]{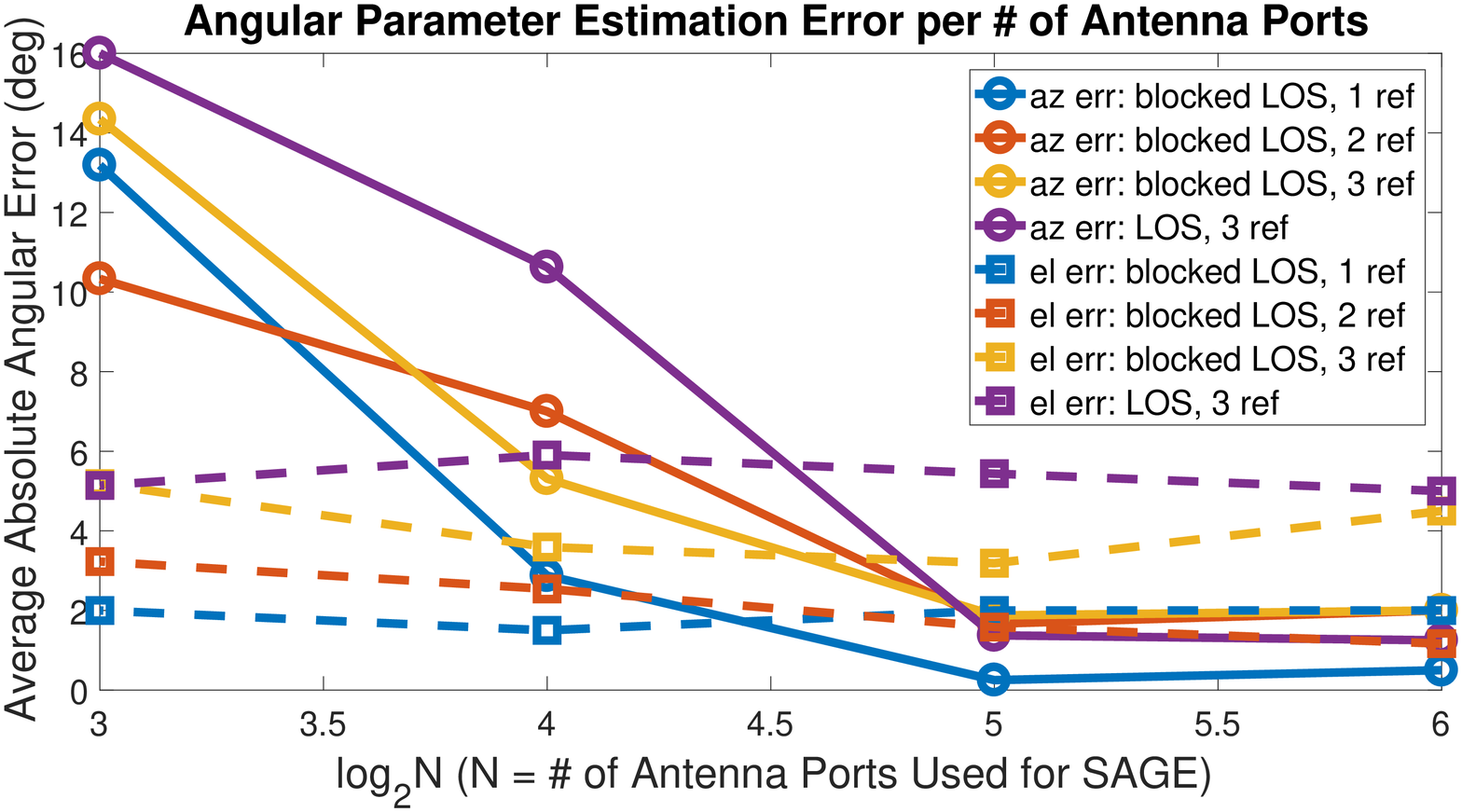}
    \centering
    \caption{Angle estimation errors depending on \# of antennas and scenarios} 
    \label{fig:angles}
    	\vspace{-1.5em}
\end{figure}

\IEEEPARstart Fig. \ref{fig:angles} shows the results of the measurement. First, the number of antennas did not have correlation with elevation estimation accuracy. The result was expected as four antennas on each column receives same signal at different elevation angles, providing accurate estimation even with just two columns. Yet, the number of antennas is expected to play a bigger role at low SNR environments outside the anechoic chamber. In contrast, the number of reflectors increased elevation estimation error by approximately one degree per additional reflector. Still, the errors were $<$ 6 degrees even for cases with three reflectors. Strong LOS may also hinder the estimation of channel parameters for multi-path components (MPCs), but not by more than 2 degrees compared to blocked LOS cases. 

\IEEEPARstart For azimuth, the average estimation error increased to up to 16 degrees as we reduced the number of antennas to 8. The beamwidth of each patch antenna is limited, and the array cannot resolve all azimuth angles if only part of 360 degrees are covered. The error also increased with the number of reflectors (except for the case with 2 reflectors) and the availability of the LOS. While not shown here, when a single row of 16 antennas covering all azimuth angles was used for SAGE estimation instead of 4 of 4x1 columns, the single row of antennas could estimate azimuth angles much more accurately compared to elevation angles. 

\IEEEPARstart Residual errors observed in the experiments may be due to the following: (i) the model mismatch (e.g., the far field assumption for the array, from both the omni-antenna and the reflectors within the chamber, or the specular reflection assumption, of the reflectors with finite extent and slightly rough surfaces), (ii) imperfect calibration of the arrays, (iii) the sensitivity of the SAGE algorithm to the above-mentioned effects, in particular when MPCs show large power differences and/or have similar parameters, and (iv) limited number of iterations and other numerical issues in applying SAGE. 

\section{Conclusion}
This study showed that the best number of antennas used during the measurement and evaluation may depend on antenna beamwidth, sounder measurement time requirement, and the channel parameter to be estimated.

\smallskip
\noindent{\bf Acknowledgement:} Part of this work was supported by NSF EECS-1731694, a gift from Samsung America, and the Belgian American Educational Foundation (B.A.E.F.). Helpful discussions with Dr. Abbasi are gratefully acknowledged.
\bibliography{mMIMO}
\bibliographystyle{IEEEtran}

\end{document}